# Spin-phonon coupling in Sr and Ti incorporated 9R-BaMnO$_3$

Bommareddy Poojitha, Anjali Rathore, and Surajit Saha*
*Department of Physics, Indian Institute of Science Education and Research, Bhopal, 462066, INDIA*
*Corresponding author: surajit@iiserb.ac.in*

**Abstract:** Materials having strong coupling between different degrees of freedom such as spin, lattice, orbital, and charge are of immense interest due to their potential device applications. Here, we have stabilized the 9R phase of BaMnO$_3$, by Sr (Ti) doping in the Ba (Mn) site using solid-state route at ambient pressure, which otherwise requires high-pressure conditions to synthesize. Crystal structure, phonon spectra, and their evolution with temperature are investigated using x-ray diffraction and Raman spectroscopic techniques. Temperature-dependent magnetization data reveal an antiferromagnetic transition at around ($T_N$ =) 262 K and 200 K for Ba$_{0.9}$Sr$_{0.1}$MnO$_3$ and BaMn$_{0.9}$Ti$_{0.1}$O$_3$, respectively. No structural phase transition or lattice instabilities are observed in the measured temperature range (80 – 400 K) for both the compounds. We have observed anomalous behaviour of four different phonon modes involving Mn/O-vibrations at low temperatures which have been attributed to spin-phonon coupling.

## Introduction:

Antiferromagnetic (AFM) materials, which were initially thought to be not so useful, have recently seen a renewed interest due to their potential for technologies such as spintronics, opto-spintronics, magnetic random-access memory (MRAM) devices, neuromorphic computing, THz information technologies, and several other applications [1-3]. AFM materials in spintronic devices are potentially advantageous over conventional ferromagnets due to their ultra-fast dynamics [2]. Interactions between spins in antiferromagnetic insulators with other degrees of freedom such as phonons play an important role in functionalization as well as spintronic memory and logic applications thus enabling antiferromagnetic spintronics [4]. Further, antiferromagnetic insulators with strong spin-phonon coupling may potentially become multiferroics where the magnetic and ferroelectric orderings can be triggered by external strain or equivalent perturbation. This mechanism was experimentally demonstrated in EuTiO$_3$ where multiferroicity arises at ~ 5 K [5], much below the room temperature. Similarly, the cubic SrMnO$_3$ has been proposed to exhibit strain-induced multiferroic behaviour originating from strong spin-phonon coupling [6]. The hexagonal SrMnO$_3$ is a near room temperature ($T_N$ = 280 K) antiferromagnet as compared to its cubic counterpart ($T_N$ =

230 K) and is more promising for room temperature applications. H. Chen *et al.*, by *ab initio* calculations, predicted that the cubic system of $Sr_{1-x}Ba_xMnO_3$ under tensile strain can be stabilized in ferroelectric-ferromagnetic state at higher Ba concentrations [7-9].

Perovskite manganites such as $AMnO_3$ with A: Ca, Sr, and Ba are found to be potential candidates with spin-phonon coupling as predicted in the literature [10,11]. Among these, $BaMnO_3$ is polymorphic, exhibiting 2H, 4H, 6H, 8H, 10H, 9R, 12R, 15R, 21R, 27R, and 33R type hexagonal structures at room temperature, where, the integer represents the number of layers while H and R denote hexagonal and rhombohedral symmetries, respectively [12,13]. The layer arrangement is very sensitive to oxygen stoichiometry which strongly depends on synthesis conditions parameterized by high-temperature and high-pressure. As the oxygen deficiency increases, 2H gradually transforms into 15, 8, 6, 10, and 4-layered ones that can be controlled at ease [13]. The crystal structure can act as an additional tuning parameter to control the coupling in this material and allow researchers to synthesize the desired phase at room temperature thus tailoring their properties. The stoichiometric 2H phase is of pure hexagonal form having a continuous chain of face-shared $MnO_6$ octahedra which is known to be multiferroic in nature with improper ferroelectricity. Ferroelectricity occurs in 2H-$BaMnO_3$ at low temperatures below the structural phase transition at $T_C$ = 130 K and it exhibits antiferromagnetic ordering below $T_N$ = 59 K [14]. Adkin *et al.*, reported antiferromagnetic transitions in the range of 250-270 K for 15R, 6H, 8H, and 10H, phases by tuning synthesis conditions [15]. On the contrary, the 9R phase is not well studied mainly due to the complexities involved in achieving a stable structure. It requires simultaneous high-pressure and high-temperature conditions to synthesize the stable 9R-phase or creation of local lattice strain achieved by chemical doping, such as, doping of Sr at Ba site and/or Ti/Bi/Ru at Mn site [15-19]. To the best of our knowledge, there are no reports on the spin-phonon coupling for the 9R phase of $BaMnO_3$ in the literature.

In this article, we have investigated the correlation between phonon and magnetic degrees of freedom in the 9R phase of $Ba_{0.9}Sr_{0.1}MnO_3$ (BSM10) and $BaMn_{0.9}Ti_{0.1}O_3$ (BMT10). The x-ray diffraction (XRD) data at room temperature confirm the 9R type hexagonal structure of BSM10 and BMT10 with space group R-3m. XRD data as a function of temperature show no discernible structural phase transition or anomaly in lattice parameters. Symmetries for the observed phonon modes are assigned according to the group theoretical predictions [20].

Detailed analysis of the temperature-dependent Raman spectra reveals that the $E_g$ mode at 344 cm$^{-1}$ associated with Mn vibrations shows anomaly with temperature whereas three other phonon modes involving Mn and O show anomaly at low temperatures below its Neel temperature ($T_N$). Mean field analysis of the phonon anomalies suggest the presence of spin-phonon interactions in BSM10 and BMT10 for these four modes and the coupling constant is found to be as high as 3.2 cm$^{-1}$ for the Mn vibration (~ 344 cm$^{-1}$) with $E_g$-symmetry.

**Synthesis and experimental details:**

Polycrystalline samples of BSM10 and BMT10 were prepared using the conventional solid-state reaction method [21]. Stoichiometric amounts of $BaCO_3$, $SrCO_3$, $MnO_2$, and $TiO_2$ (Sigma Aldrich) were taken as the precursors. Stoichiometric mixture for BSM10 was ground well for 3 hours and calcined at 800°C, 900°C, and 1000°C for 24 hours each with intermediate grinding. Resultant powder is ground again for about 3 hours, pressed into a pellet of 10 mm and sintered at 1200°C for 12 hours. Similarly, For BMT10, two calcinations were done at 1250°C and 1300°C followed by the sintering at 1400°C for 2 hours each with intermediate grindings [22]. PANalytical Empyrean x-ray diffractometer with Cu-K$_\alpha$ radiation of wavelength 1.5406 Å attached with Anton paar TTK 450 heating stage was used for temperature-dependent powder x-ray diffraction (PXRD) measurements. Raman spectra were collected in the backscattering configuration using a LabRAM HR Evolution Raman spectrometer attached to a 532 nm laser excitation source and Peltier cooled charge coupled device (CCD) detector. HFS600E-PB5 Linkam stage was used for temperature-dependent Raman measurements. DC Magnetization measurements were carried out using Quantum Design SQUID-VSM (Superconducting Quantum Interference Device with Vibrating Sample Magnetometer).

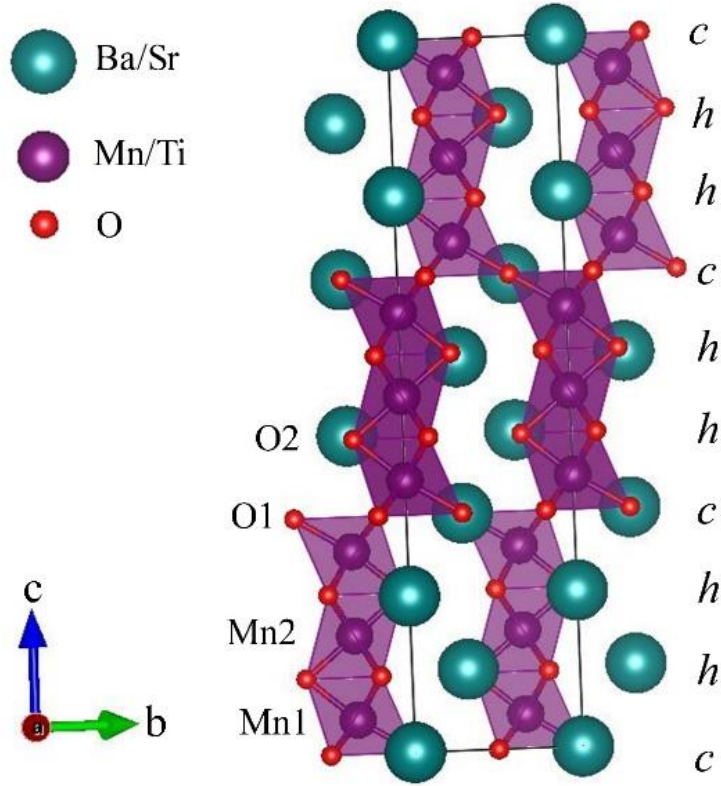

*Figure 1. (Color online) Crystal structure (unit cell) of Sr or Ti doped 9R-BaMnO$_3$.*

## Results and Discussion:

The 9R-phase of BaMnO$_3$ contains nine layers of BaO$_3$ per unit cell along the hexagonal c-axis with *"ababcbcac"* stacking, identical to that in Samarium metal [23]. This structure can also be visualized as a combination of cubic (c) and hexagonal (h) layers with the stacking sequence (chh)$_3$ within the unit cell. In other words, the unit cell is built of triplets of face-shared MnO$_6$ octahedra forming Mn$_3$O$_{12}$ trioctahedron unit connected to another trioctahedron unit through corner-sharing. Crystal symmetry is rhombohedral with the space group R-3m (No. 166) and point group symmetry $D_{3d}^5$ having three molecules per unit cell as shown in Figure 1. In the 9R phase of BaMnO$_3$, both the O and Mn atoms have two irreducible sites. The Mn and O atoms belonging to an octahedron that shares corner as well as face with other octahedra are labelled as Mn1 and O1 whereas those associated with a purely face-shared octahedron are called as Mn2 and O2 [24].

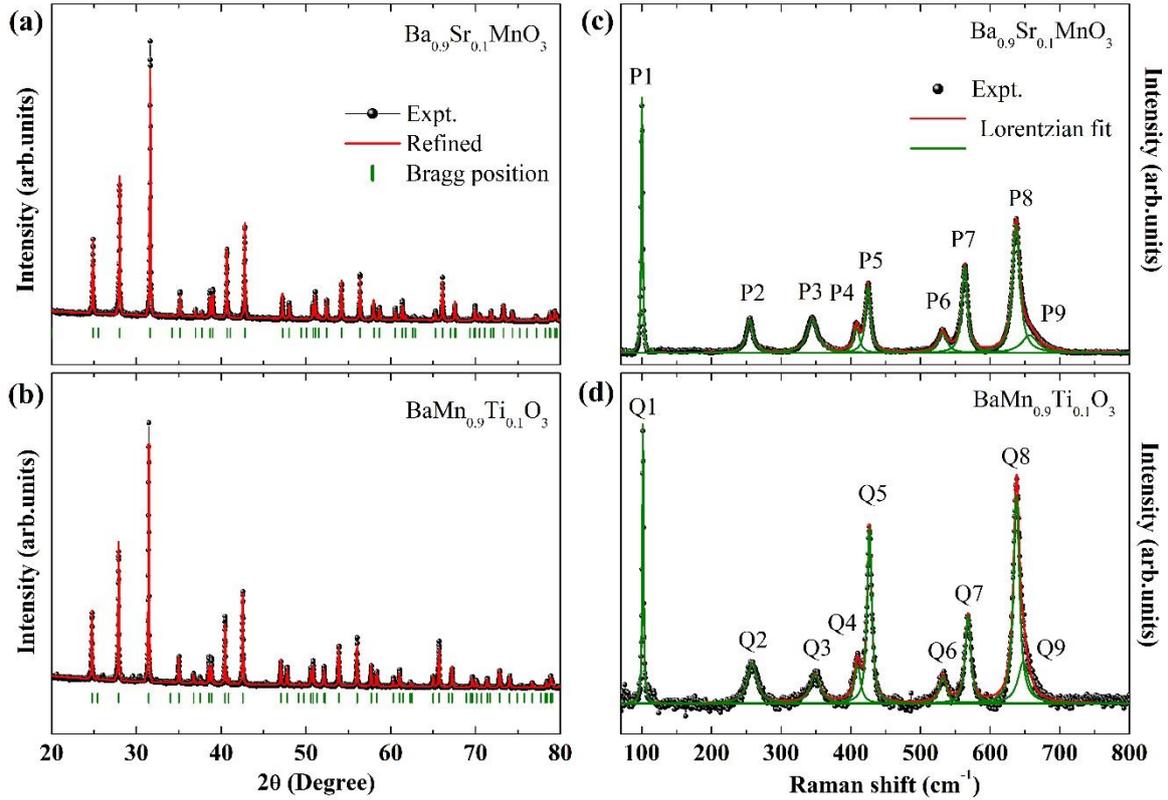

*Figure 2. (Color online) (a, b) X-ray diffraction pattern, and (c, d) Raman spectra at room temperature for $Ba_{0.9}Sr_{0.1}MnO_3$ and $BaMn_{0.9}Ti_{0.1}O_3$, respectively.*

We have synthesized the 9R phase of $BaMnO_3$ at ambient pressure by incorporating Sr (Ti) at Ba (Mn) site. The crystal structure is confirmed by Rietveld analysis of x-ray diffraction patterns, shown in Figure 2 (a, b). The refined unit cell parameters at room temperature are a = b = 5.6426 Å and c = 20.876 Å for BSM10, whereas a = b = 5.6728 Å and c = 20.955 Å for BMT10, thus matching well with the report on $BaMnO_3$ [15]. The ionic radii of $Ba^{+2}$, $Sr^{+2}$, $Mn^{+4}$, and $Ti^{+4}$ are 1.35, 1.18, 0.53, and 0.61 Å, respectively [25]. Therefore, substitution of $Ba^{+2}$ by $Sr^{+2}$ compresses the unit cell whereas replacement of $Mn^{+4}$ by $Ti^{+4}$ leads to expansion of the lattice. The Mn-Mn separation in the face-shared octahedra is 2.4696 Å and 2.4790 Å in BSM10 and BMT10, respectively, which also confirms that the dopants are incorporated at the respective sites. The O(2)-O(2) distance (2.5166 Å and 2.5301 Å) in the face-shared octahedra is shorter than the O(1)-O(1) (2.8213 Å and 2.8364 Å) as well as O(1)-O(2) (2.7410 Å and 2.7526 Å) distances (for BSM10 and BMT10, respectively) in the face/corner-shared octahedra which is an important feature of the 9R type perovskites [24,26]. Optical phonon modes at the zone centre predicted by group theory for 9R phase can be expressed with irreducible representations as follows [20]:

$$\Gamma = 4A_{1g}(R) + 5E_g(R) + 7A_{2u}(IR) + 9E_u(IR) + 2A_{1u}(S) + A_{2g}(S)$$

where, R, IR, and S represent Raman active, Infrared active, and silent modes, respectively.

*Table. I:* *Assignment of the Raman active phonons of $Ba_{0.9}Sr_{0.1}MnO_3$ and $BaMn_{0.9}Ti_{0.1}O_3$ at room temperature.*

| Mode Symmetry | $Ba_{0.9}Sr_{0.1}MnO_3$ $\omega$ (cm$^{-1}$) | $BaMn_{0.9}Ti_{0.1}O_3$ $\omega$ (cm$^{-1}$) | Atoms involved in the vibration |
|---|---|---|---|
| $E_g$ | 100 (P1) | 101 (Q1) | Ba/Sr |
| $A_{1g}$ | 255 (P2) | 258 (Q2) | Ba/Sr |
| $E_g$ | 344 (P3) | 348 (Q3) | Mn/Ti |
| $A_{1g}$ | 408 (P4) | 409 (Q4) | Mn/Ti |
| $E_g$ | 424 (P5) | 426 (Q5) | O |
| $E_g$ | 531 (P6) | 532 (Q6) | O |
| $A_{1g}$ | 563 (P7) | 568 (Q7) | O |
| $A_{1g}$ | 637 (P8) | 638 (Q8) | O |
| $E_g$ | 656 (P9) | 652 (Q9) | O |

The Raman spectra of BSM10 and BMT10 at room temperature are shown in Figure 2 (c, d). Nine phonon modes are observed at room temperature (P1 to P9 for BSM10 and Q1 to Q9 for BMT10) that match well with the reported spectra of isostructural compounds $BaRuO_3$ [27] and $BaMn_{0.85}Ti_{0.15}O_3$ [18]. Phonon symmetries, as listed in Table I, are assigned based on the report on 9R-$BaRuO_3$ [27]. The vibration of O(2) gives rise to IR active modes while O(1) takes part in both IR and Raman active phonons. On the other hand, the Mn(1) atoms participate in the Raman active vibrations. Since Raman spectra are discussed in the present report, henceforth the O(1) and Mn(1) are abbreviated as O and Mn for convenience. The modes P1/Q1 ($E_g$) and P2/Q2 ($A_{1g}$) arise due to Ba (Sr) vibrations, whereas, the P3/Q3 ($E_g$) and P4/Q4 ($A_{1g}$) modes originate from Mn (Ti) vibrations. The remaining $3E_g$ and $2A_{1g}$ (P5 to P9 and Q5 to Q9) modes arise from oxygen vibrations [27]. Raman spectra of BSM10 and BMT10 at a few selected temperatures are shown in Figures 3(a) and 4(a), respectively. As no new phonon mode appears or existing phonon disappears in the measured temperature range of 80 to 400 K, it is an indicative of the absence of structural phase transition as also corroborated by temperature-dependent XRD, discussed later. Phonon parameters at each temperature are extracted by fitting the spectra with the Lorentzian function. The phonon frequencies as a function of temperature for BSM10 and BMT10, respectively, are shown in Figures 3(b) and 4(b). The frequencies of all the phonon modes show a red-shift with increasing temperature. However, the mode P3/Q3 at 344/348 cm$^{-1}$ ($E_g$), which is related to Mn (and Ti) vibrations, shows an anomalous behaviour at low temperatures, i.e. at low temperatures the frequency increases with increasing temperature.

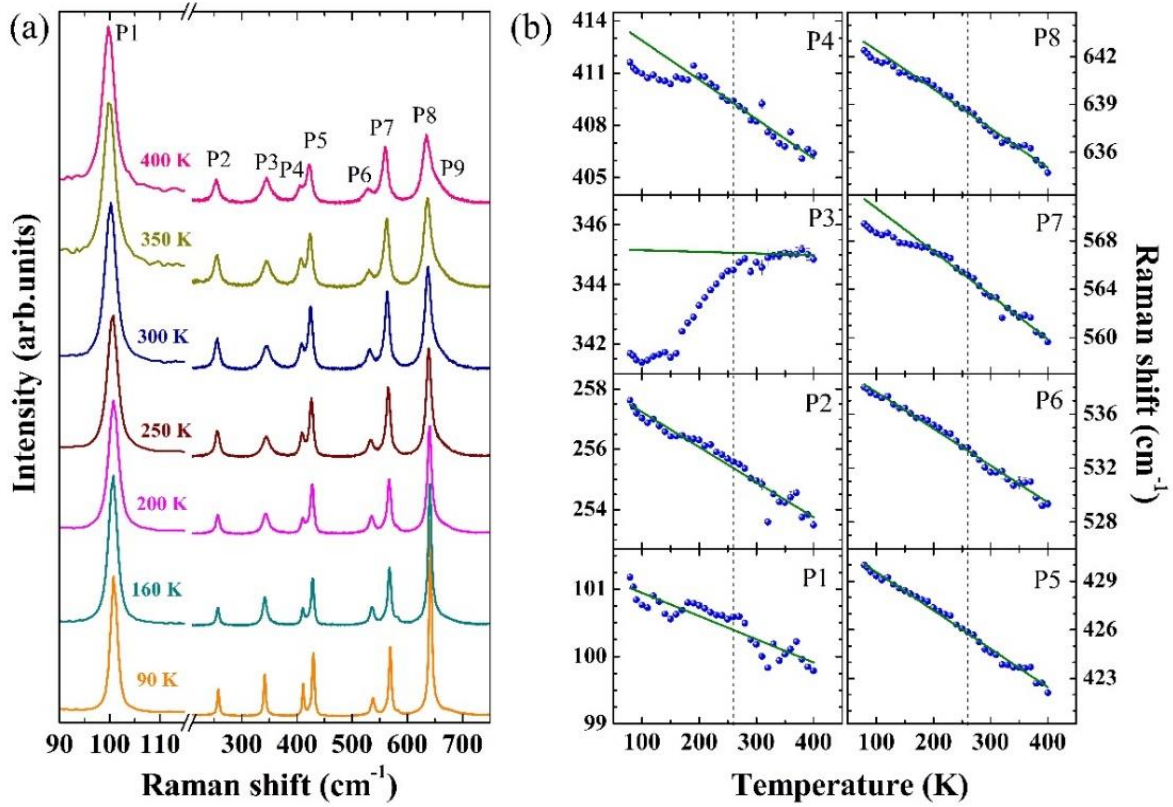

*Figure 3.* (Color online) (a) Raman spectra of Ba$_{0.9}$Sr$_{0.1}$MnO$_3$ at a few temperatures and (b) phonon frequencies as a function of temperature. Anomalies can be seen for the modes P3, P4, P7, and P8 at low temperatures (bounded below the dashed line). Error bars in (b) are smaller than symbol size.

In general, the temperature-dependent phonon frequency can be expressed as [28]:

$$\omega(T) = \omega_{anh}(T) + \Delta\omega_{el-ph}(T) + \Delta\omega_{sp-ph}(T) \qquad (1)$$

The first term $\omega_{anh}(T)$ accounts for the contribution purely due to anharmonicity. The term $\Delta\omega_{el-ph}(T)$ is the change in phonon frequency arising due to electron-phonon coupling which can be considered to be absent in the insulating BSM10 and BMT10. $\Delta\omega_{sp-ph}(T)$ is the effect of renormalization of the frequency due to spin-phonon coupling. The shift in the phonon frequency due to anharmonicity can be simplified by considering the three phonon process (cubic anharmonicity) and neglecting higher order processes which can be expressed as [28,29]:

$$\omega_{anh}(T) = \omega_0 + A\left[1 + \frac{2}{\left(e^{\frac{\hbar\omega_0}{2k_BT}}-1\right)}\right] \quad (2)$$

Similarly, the temperature dependence of phonon linewidth considering the cubic anharmonicity can be written as:

$$\Gamma_{anh}(T) = \Gamma_0 + C\left[1 + \frac{2}{\left(e^{\frac{\hbar\omega_0}{2k_BT}}-1\right)}\right] \quad (3)$$

where, $\omega_0$ and $\Gamma_0$ are frequency and linewidth of the phonon at absolute zero temperature, $A$ and $C$ are cubic anharmonic coefficients for frequency and linewidth, respectively, $\hbar$ is reduced plank constant, $k_B$ is Boltzmann constant and T is the variable temperature.

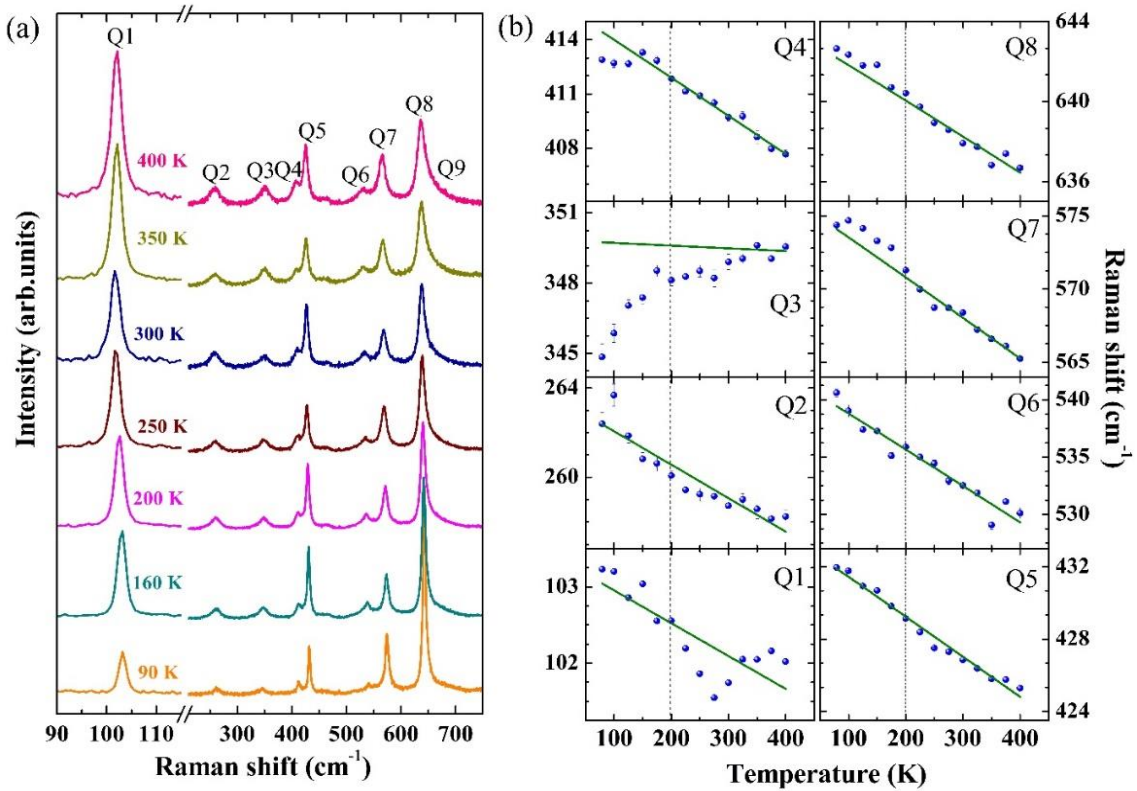

*Figure 4. (Color online) (a) Raman spectra of BaMn$_{0.9}$Ti$_{0.1}$O$_3$ at a few temperatures and (b) phonon frequencies (with error bar) as a function of temperature. Clear anomalies are seen for the modes Q3 and Q4, while an appreciable deviation is observed for Q7 and Q8 at low temperatures (bounded below the dashed line).*

Solid lines in Figures 3b and 4b represent the fitting for experimental data with Eq. 2. It is to be noted that the frequency of the mode P3 decreases swiftly with lowering temperature below ~ 262 K down to the lowest measured temperature (~ 80 K) thus deviating from the expected anharmonic behaviour (solid line). In addition, P4, P7, and P8 modes also show deviation from the conventional anharmonic behaviour in the low temperature region. The origin of the phonon anomalies observed below 262 K will be discussed later. In comparison, similar phonon anomalies are seen for the modes Q3 and Q4 in BMT10 below ~ 200 K (Figure 4b). In order to verify if any structural transition or deformation plays a role in the observed phonon anomalies at low temperatures, we have performed temperature-dependent XRD measurements in the temperature range of 90 to 400 K. Figure 5 (a, b) shows the lattice parameters of BSM10 and BMT10 as a function of temperature revealing a thermal expansion of the lattice. Rietveld refinement of their temperature-dependent diffraction patterns reveals that there is no structural phase transition or measurable lattice deformation down to 90 K.

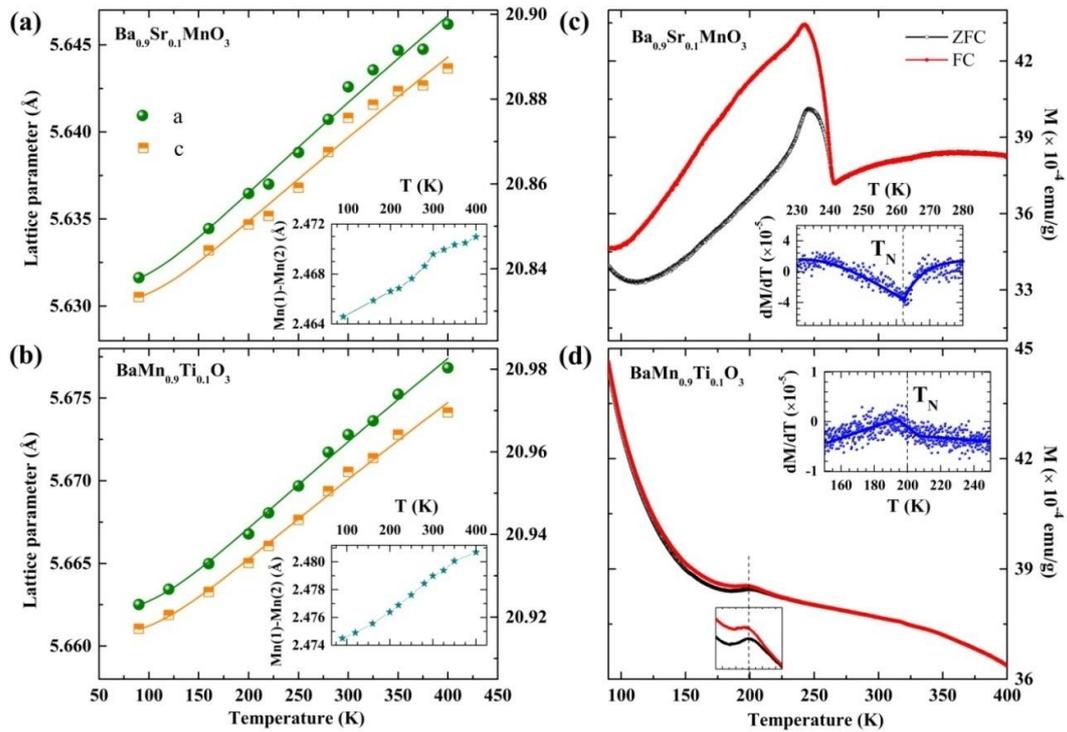

*Figure 5*. *(Color online) (a, b) Lattice parameters and (c, d) magnetization (M = magnetic moment per unit gram) of $Ba_{0.9}Sr_{0.1}MnO_3$ and $BaMn_{0.9}Ti_{0.1}O_3$ as a function of temperature (T). Solid lines in (a), (b) are fit to the lattice parameters with equations discussed in the text (Eq. 4). Insets in (a) and (b) show the Mn(1)-Mn(2) distance as a function of temperature while the insets in (c) and (d) show the temperature-derivative of magnetization representing magnetic transition temperatures. The solid lines in the insets are guide to eye. Bottom inset of (d) shows the enlarged view of M-T around the magnetic transition temperature.*

Thermal expansion of BSM10 and BMT10 are unremarkable and display conventional positive thermal expansion. Mn-Mn distance also shows a systematic increment upon heating (insets of Figure 5 (a, b)). The temperature-dependent lattice parameters can be fitted by [30]:

$$a(T) = a_0 \left[1 + \frac{be^{\frac{d}{T}}}{T(e^{\frac{d}{T}}-1)^2}\right] \quad \text{and} \quad c(T) = c_0 \left[1 + \frac{fe^{\frac{g}{T}}}{T(e^{\frac{g}{T}}-1)^2}\right] \quad (4)$$

where, $a_0$ and $c_0$ are the in-plane and out-of-plane lattice constants at 0 K, whereas b, d, f, and g are fitting parameters. The obtained values are $a_0$ = 5.6311 Å, b = 1.95 K, and d = 499 K for BSM10 and $a_0$ = 5.6621 Å, b = 2.26 K, and d = 538 K for BMT10, respectively. Similarly, $c_0$ = 20.832 Å, f = 2.21 K, and g = 524 K for BSM10 and $c_0$ = 20.916 Å, f = 2.27 K, and g = 540 K for BMT10, respectively. Unit cell parameters and the bond lengths show no signatures of structural transition or local lattice deformation in the investigated temperature range. Therefore, the role of structural transition/ deformation in the observed phonon anomalies below 262 and 200 K in BSM10 and BMT10, respectively, can be completely ruled out. In order to verify the possible magnetic ordering as reported in similar manganite systems [16,31,32] and its role in the phonon anomalies, magnetization measurements were performed as a function of temperature in both zero field cooling (ZFC) and field cooling (FC) cycles under the applied magnetic field of 500 Oe, as shown in Figure 5 (c, d). The magnetization (M = magnetic moment per unit gram) is found to be weakly temperature (T)-dependent in the high temperature region for both BSM10 and BMT10. Upon cooling, presence of a peak in magnetization and a bifurcation in ZFC and FC curves are observed signifying a magnetic transition from paramagnetic to antiferromagnetic phase at low temperatures. The Neel temperature ($T_N$) for BSM10 and BMT10 are estimated to be ~ 262 K and ~ 200 K, respectively, by plotting the temperature derivative of magnetization as a function of temperature (see insets of Figure 5(c, d)). The contrasting difference between the M-T data of BSM10 and BMT10 may be attributed to the dilution of $Mn^{+4}$ moments by $Ti^{+4}$ in BMT10. A relatively higher net magnetization in BMT10 at low temperatures as opposed to BSM10 may be due to the unpaired $Mn^{+4}$ ions in BMT10 arising from the replacement of $Mn^{+4}$ by $Ti^{+4}$. A detailed study on the magnetic properties of BSM10 and BMT10 is out of the scope of this article and hence will be reported elsewhere.

**Signatures of spin-phonon coupling:**

To gain an insight into the origin of the phonon anomalies around and below $T_N$, we have carefully fitted the ω vs T data above $T_N$ with anharmonicity (Eq. 2) and extrapolated till the lowest measured temperature (80 K). A clear deviation from an expected anharmonic behaviour can be seen for the modes P3, P4, P7, P8, Q3, and Q4, as shown in Figures 3b and 4b. Renormalization of a phonon associated with magnetic ordering may arise due to magnetostriction effect and/or spin-phonon coupling. Magnetostriction contributes to the phonon frequency through a change in the lattice volume or bond lengths caused by magnetic ordering below $T_N$. Since our temperature-dependent XRD results (Figure 5 a, b) show no signatures of structural transition or measurable lattice deformation in the studied temperature range for both BSM10 and BMT10, the possibility of magnetostriction can be safely ruled out.

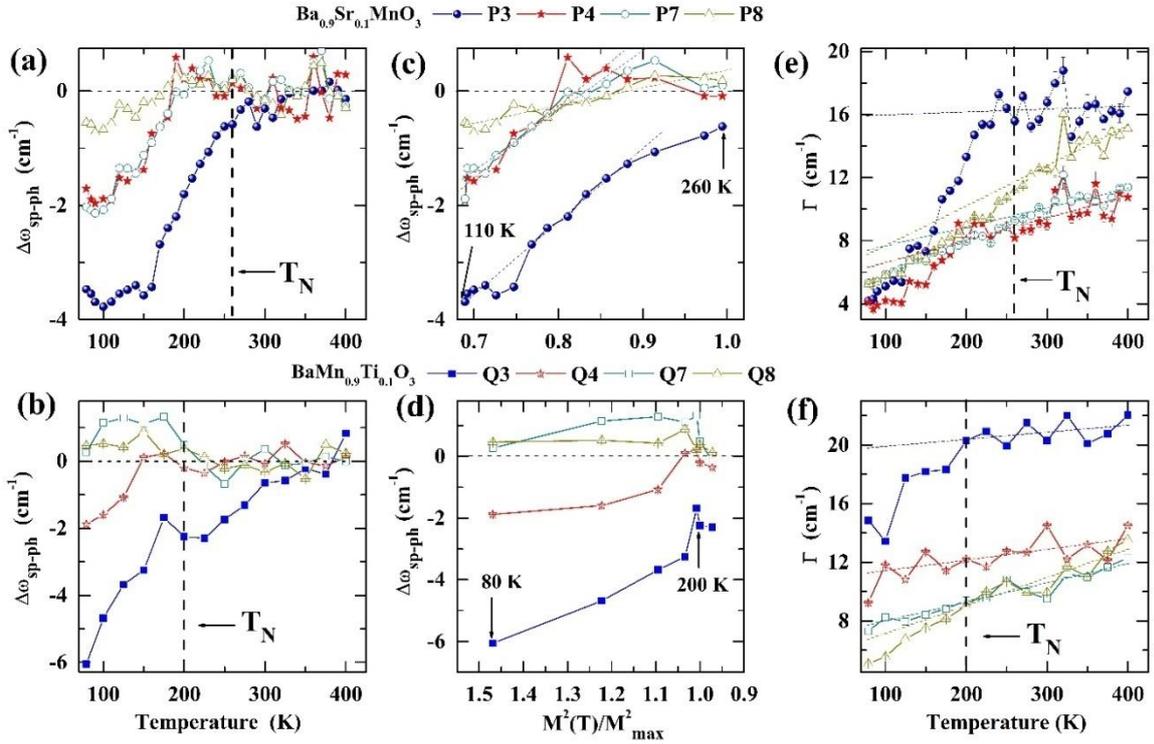

*Figure 6. (Color online) (a, b) The deviation of phonon frequency from anharmonic behaviour $(\Delta\omega_{sp-ph})$ as a function of temperature (see text). (c, d) The variation of $\Delta\omega_{sp-ph}(T)$ with $\left[\frac{M(T)}{M_{max}}\right]^2$, and (e, f) temperature-dependent phonon linewidth. The dashed lines (in e and f) are the anharmonic fits to the linewidths (Eq. 3) showing a deviation below the para-to-antiferromagnetic transition temperature ($T_N$).*

We, therefore, attribute the phonon anomalies and deviations from anharmonic behaviour (in P3, P4, P7, P8, Q3, and Q4) to spin-phonon coupling. To recall, the 9R phase of $BaMnO_3$ is a stack of nine-layers of octahedral units in a cubic and hexagonal layer arrangement with (chh)$_3$ stacking sequence (see Figure 1) where the Raman active modes involve vibrations of Mn and O atoms belonging to the corner-shared octahedra. Further, the $Mn^{4+}$ ions in the corner-shared octahedra interact antiferromagnetically via superexchange mechanism [15]. Therefore, a spin-phonon coupling can arise owing a modulation of the superexchange integral due to the associated Mn/O atomic vibrations [33,34]. Within the mean-field approximation, renormalization of a phonon frequency due to the effect of magnetic ordering below $T_N$ can be expressed as [33]:

$$\omega(T) = \omega_{anh}(T) + \lambda <S_i.S_j>$$

where, $\omega_{anh}(T)$ is the frequency purely due to phonon anharmonicity (given in Eq. 2) in the absence of spin-spin correlations. The term $<S_i.S_j>$ is related to the spin-spin correlation function between two nearest neighbour spins localized at any $i^{th}$ and $j^{th}$ sites, and $\lambda$ is the spin-phonon coupling coefficient. In presence of spin-phonon coupling, the phonon frequency deviates from anharmonic behaviour (Figures 3b and 4b) and therefore, $\Delta\omega_{sp-ph}(T) = \omega(T) - \omega_{anh}(T)$ gives the measure of the deviation. Figure 6 (a, b) shows the deviation for the modes P3, P4, P7, P8, Q3, Q4, Q7, and Q8 as a function of temperature. We have estimated the strength of spin-phonon coupling using the relation [34-36]:

$$\Delta\omega_{sp-ph}(T) = \lambda <S_i.S_j> \approx n\lambda \left[\frac{M(T)}{M_{max}}\right]^2 \quad (5)$$

Here, $n$ is the in-plane coordination number for magnetic ion and $M(T)$ is the temperature-dependent magnetization with $M_{max}$ being the maximum magnetization. The deviation, $\Delta\omega_{sp-ph}(T)$, as a function of $\left[\frac{M(T)}{M_{max}}\right]^2$ is nearly linear, as shown in Figure 6 (c, d). The slope of $\Delta\omega_{sp-ph}(T)$ $vs$ $\left[\frac{M(T)}{M_{max}}\right]^2$ curve gives the estimate of the spin-phonon coupling strength ($\lambda$) for the phonon modes which are listed in Table II.

***Table II:*** *Spin-phonon coupling constants for four different phonon modes of $Ba_{0.9}Sr_{0.1}MnO_3$ and $BaMn_{0.9}Ti_{0.1}O_3$.*

| Phonon | λ (cm⁻¹) |
|---|---|
| P3 | 3.2 |
| P4 | 3.1 |
| P7 | 2.7 |
| P8 | 0.9 |
| Q3 | 3.0 |
| Q4 | 2.0 |
| Q7 | 0.3 |
| Q8 | 0.1 |

As evident from the above observations, the modes P3 and P4 in BSM10 as well as Q3 and Q4 in BMT10 have a reasonably strong spin-phonon coupling ($\lambda \sim 3.2$ cm$^{-1}$), in comparison to the reports on manganite (4H) $Sr_{0.6}Ba_{0.4}MnO_3$ ($\lambda \sim 2.2$ cm$^{-1}$[31]), $Sr_2CoO_4$ ($\lambda \sim 3.5$ cm$^{-1}$ [36]), NiO ($\lambda \sim -7.9$ cm$^{-1}$ and 14.1 cm$^{-1}$ for TO and LO phonons, respectively [37]), $MnF_2$ ($\lambda \sim 0.4$ cm$^{-1}$[38]), $FeF_2$ ($\lambda \sim 1.3$ cm$^{-1}$[38]), $ZnCr_2O_4$ ($\lambda \sim 3.2$ to 6.2 cm$^{-1}$[39]), $NaOsO_3$ ($\lambda \sim 40$ cm$^{-1}$[40]), CuO ($\lambda \sim 50$ cm$^{-1}$[41]), $La_2CoMnO_6$ ($\lambda \sim 1.7$ to 2.1 cm$^{-1}$ [42], $Pr_2CoMnO_6$ ($\lambda \sim 0.51$ to 1.61 cm$^{-1}$ [43, 44]), and $Cr_2Ge_2Te_6$ ($\lambda \sim 0.24$ to 1.2 cm$^{-1}$ [45]). On the other hand, the modes Q7 and Q8 in BMT10 have low spin-phonon coupling as opposed to the modes P7 and P8 in BSM10 which may be because of the dilution of magnetic ions in BMT10 due to the replacement of $Mn^{+4}$ by $Ti^{+4}$. In addition to the changes in the phonon frequencies due to spin-phonon coupling, we have also observed deviations in the phonon linewidth ($\Gamma$) from the expected anharmonic behaviour (Eq. 3) below $T_N$, as shown in Figure 6 (e, f). Phonon renormalization due to spin-phonon coupling is expected to lead to a change in the phonon lifetime which is associated with the phonon linewidth ($\Gamma \propto 1/\tau$). Therefore, the deviation in linewidth from anharmonic behaviour further signifies the role of spin-phonon coupling in the Sr/Ti incorporated 9R-BaMnO$_3$. Further detail on the phonon linewidths is given in the supplementary material (Figures S1 and S2).

## Conclusion:

In conclusion, we have synthesized the 9R phase of BaMnO$_3$ with Sr and Ti incorporation at the Ba and Mn sites, respectively, by solid-state reaction method. The doping promotes stabilization of the 9R phase by creating a local stress arising due to the differences in the ionic radii thus making the synthesis easier as opposed to the complexities involved in the high-pressure synthesis of this polymorph [16]. Temperature-dependent XRD and Raman spectroscopic results reveal no structural transitions or lattice deformations in the investigated temperature range. Four different phonon modes with $E_g$ and $A_{1g}$ symmetries involving Mn and O vibrations have been identified exhibiting high spin-phonon coupling leading to an anomalous behaviour below the para- to antiferro-magnetic transition temperature ($T_N$). Antiferromagnetic materials with high spin-phonon coupling are of great interest for technological applications and we believe that the 9R-BaMnO$_3$ phase can be a good candidate.

**Acknowledgement:**

Authors acknowledge IISER Bhopal for research facilities, B. P. acknowledges the University Grant Commission for fellowship and S. S. acknowledges DST/SERB for research funding (project No. ECR/2016/001376).

22. Phan T L, Zhang P, Grinting D, Yu S C, Nghia N X, Dang N V, and Lam V D, Influences of annealing temperature on structural characterization and magnetic properties of Mn-doped $BaTiO_3$ ceramics, J. Appl. Phys. **112** (2012) 013909, https://doi.org/10.1063/1.4733691.

23. Ellinger F H and Zachariasen W H, The Crystal Structure of Samarium Metal and of Samarium Monoxide, J. Am. Chem. Soc. **75** (1953) 5650-5652, https://doi.org/10.1021/ja01118a052.

24. Richard J. D. Tilley, Perovskites, Structure-Property Relationships, first ed., Wiley, UK, 2016.

25. Shannon R D, Revised effective ionic radii and systematic studies of interatomic distances in halides and chalcogenides, Acta Crystallographica. **A32** (1976) 751-767, https://doi.org/10.1107/S0567739476001551.

26. Yusa H, Sata N, and Ohishi Y, Rhombohedral (9R) and hexagonal (6H) perovskites in barium silicates under high pressure, American Minaralogist, **92** (2007) 648-654, https://doi.org/10.2138/am.2007.2314.

27. Lee Y S, Noh T W, Park J H, Lee K B, Cao G, Crow J E, Lee M K, Eom C B, Oh E J and Yang I-S, Temperature-dependent Raman spectroscopy in $BaRuO_3$ systems, Phys. Rev. B 65 (2002) 235113, https://doi.org/10.1103/PhysRevB.65.235113.

28. Balkanski M, Wallis R F, and Haro E, Anharmonic effects in light scattering due to optical phonons in silicon, Phys. Rev. B **28** (1983) 1928-1934, https://doi.org/10.1103/PhysRevB.28.1928.

29. Klemens P G, Anharmonic Decay of Optical Phonons, Physical Review 148 (1966) 845-848, https://doi.org/10.1103/PhysRev.148.845.

30. Charles Kittel, Introduction to Solid State Physics, seventh ed., Wiley, New York, 2003.

31. Rawat R, Phase D M, and Choudhary R J, Spin-phonon coupling in hexagonal $Sr_{0.6}Ba_{0.4}MnO_3$, J. Magn. Magn. Mater. **441** (2017) 398-403, https://doi.org/10.1016/j.jmmm.2017.05.089.

32. Fang Y, Mingtao L, and Chen F, Observation of Magnetic Phase Transition and Magnetocaloric Effect in $Ba_{1-x}Sr_xMnO_{3-\delta}$, J. Supercond. Nov. Magn. **31** (2018) 3787, https://doi.org/10.1007/s10948-018-4648-1.

33. Lockwood D J, Spin-Phonon interaction and mode softening in $NiF_2$, Low Temp. Phys. **28** (2002) 505-509, https://doi.org/10.1063/1.1496657.
16

# Supplementary material

# Spin-phonon coupling in Sr and Ti incorporated 9R-BaMnO$_3$

Bommareddy Poojitha, Anjali Rathore,and Surajit Saha*

*Department of Physics, Indian Institute of Science Education and Research, Bhopal, 462066, INDIA*

*Corresponding author:surajit@iiserb.ac.in*

This supplementary material contains additional data and detailed information about fittings performed during data analysis and explanation for the theoretical models used by the authors to establish the spin phonon coupling in the studied systems. The figures and corresponding descriptions are given below.

## S1. Anharmonic model/ phonon anharmonicity

The shift in the phonon frequency ($\omega$) and linewidth ($\Gamma$) due to anharmonicity are given by Eq. 2 and 3 in the main text. Figure S1 (a, b) shows a schematic of the temperature-dependent phonon frequency and linewidth due to anharmonicity. It can be seen clearly that the frequency decreases linearly with increasing temperature in the entire temperature range except at very low temperatures where the frequency tends to saturate (because at very low temperatures the phonon anharmonicity is absent or negligibly small). Similarly, the phonon linewidth is expected to increase linearly with the temperature due to anharmonic phonon-phonon interactions. Therefore, both $\omega$ and $\Gamma$ follow linear behaviour with temperature in the investigated temperature range (80-400 K), as shown schematically in the Figure S1(c, d) (for the P1 mode near 100 cm$^{-1}$). We have fitted the frequency and linewidth of the phonon modes (Figures 3b, 4b, and S2) using Eq. 2 and 3, respectively. This anharmonic model gives an excellent description for both the frequency and linewidth of phonon modes P1, P2, P5, P6, Q1, Q2, Q5, and Q6, as shown in Figures 3b, 4b, and S2. The corresponding fitting parameters are listed in Table-S1.

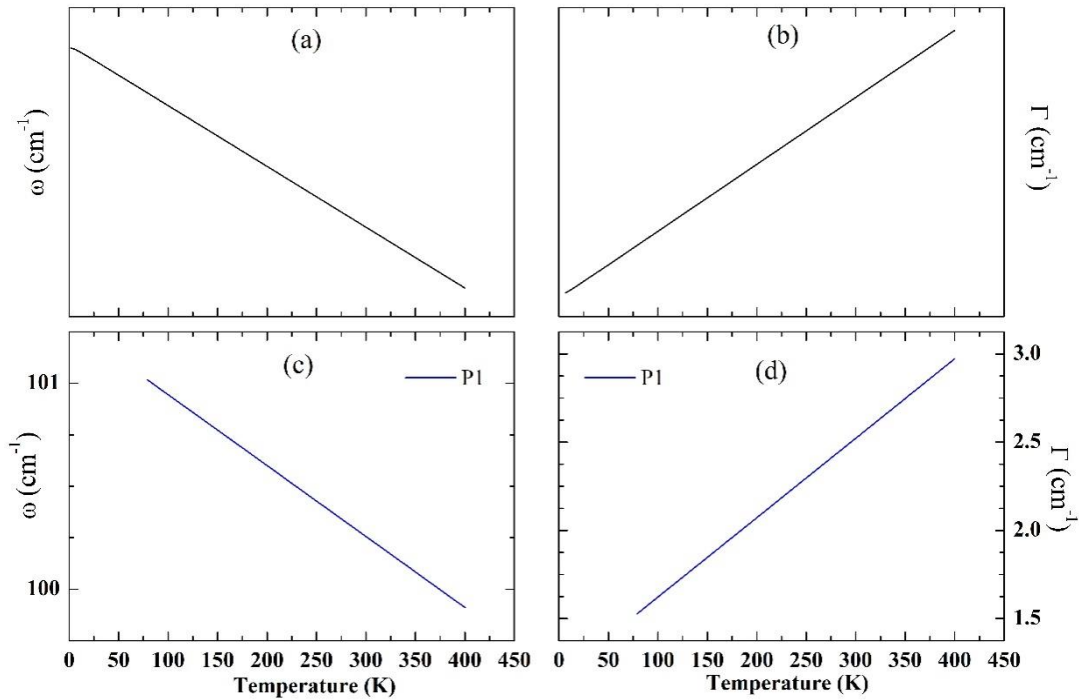

*Figure S1: (a, b) Schematics for the change in phonon frequency and linewidth due to anharmonicity as a function of temperature. (c, d) Simulated anharmonic curves for frequency and linewidth for the mode P1(of BSM10) in the temperature range of 80-400 K.*

However, the phonon modes P3 and Q3 do not follow the expected anharmonic trend (shown by green lines in Figures 3b and 4b), instead show highly anomalous behaviour with temperature (i.e. an increase in frequency with increasing temperature). In addition to that, the phonon modes P4, P7, P8, Q4, Q7, and Q8 also show anomalies below the magnetic transition temperature by deviating from the anharmonic trends (Figures 3b and 4b in the manuscript).

*Table-S1: List of fitting parameters with anharmonic model for various phonons in $Ba_{0.9}Sr_{0.1}MnO_3$ and $BaMn_{0.9}Ti_{0.1}O_3$.*

| Phonon | $\omega_0$ | $A$ | $\Gamma_0$ | $C$ |
|---|---|---|---|---|
| P1 | 101.3 ± 0.1 | -0.020 ± 0.001 | 1.17 ± 0.06 | 0.026 ± 0.001 |
| P2 | 258.4 ± 0.1 | -0.172 ± 0.007 | 2.28 ± 0.45 | 0.435 ± 0.026 |
| P5 | 431.9 ± 0.1 | -0.588 ± 0.010 | 3.40 ± 0.14 | 0.441 ± 0.013 |
| P6 | 540.5 ± 0.1 | -0.852 ± 0.017 | 3.61 ± 1.19 | 1.080 ± 0.144 |
| Q1 | 103.4 ± 0.2 | -0.025 ± 0.005 | 1.41 ± 0.08 | 0.025 ± 0.002 |
| Q2 | 263.5 ± 0.4 | -0.226 ± 0.026 | 15.06 ± 0.34 | 0.206 ± 0.020 |
| Q5 | 433.7 ± 0.2 | -0.552 ± 0.022 | 3.39 ± 0.29 | 0.382 ± 0.027 |
| Q6 | 542.1 ± 0.6 | -0.991 ± 0.073 | 8.30 ± 2.02 | 0.603 ± 0.241 |

**S2. Deviation from the anharmonic model**

To get more insight into the observed anomalies, we have carefully analysed the data and extracted the deviation from the expected anharmonic phonon frequency. In order to quantify the deviation from the anharmonic behaviour, we have fitted the modes P3, P4, P7, P8, Q3, Q4, Q7, and Q8 with anharmonic equation (Eq. 2) above $T_N$ and extrapolated (simulated) the anharmonic trend to the lowest temperature measured (80 K). The deviation in the phonon frequency from the anharmonic behaviour is calculated by the equation;

$$\Delta\omega(T) = \omega(T) - \omega_{anh}(T)$$

where, $\omega(T)$ is the experimentally observed phonon frequency at a temperature T (Figure S3).

To further confirm the presence of spin-phonon coupling, we have also analysed the temperature-dependent phonon linewidth for all the phonon modes. The linewidth of the modes P1, P2, P5, P6, Q1, Q2, Q5, and Q6 follow the expected anharmonic trend (see Figure S2). However, the linewidth of P3, P4, P7, P8, Q3, Q4, Q7, and Q8 phonons decrease distinctly below $T_N$ showing the deviation from the effect of anharmonic phonon-phonon interactions (see Figure 6(e,f)). Linewidth is a measure of the phonon life-time ($\tau \propto 1/\Gamma$). On a similar note to the temperature dependence of frequency (eq. 1), the linewidth also changes due to phonon anharmonicity, electron-phonon interaction, and spin-phonon coupling [*Phys. Rev. B* **28** (1983) 1928-1934]. We can safely rule out the possibility of electron-phonon coupling in BSM10 and BMT10 as they are insulators. So, the observed anomalies in frequency and linewidth of phonons are attributed to spin-phonon coupling [*Europhysics Letters* **101** (2013) 17008].

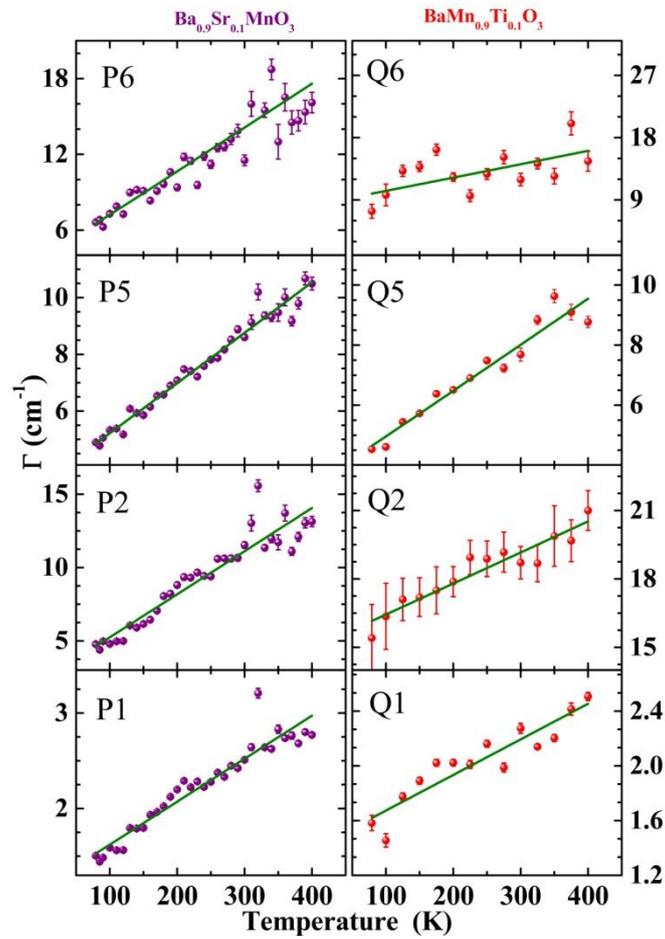

*Figure S2: Phonon linewidth as a function of temperature for anharmonic modes in $Ba_{0.9}Sr_{0.1}MnO_3$ and $BaMn_{0.9}Ti_{0.1}O_3$. Green solid lines represent the anharmonic fit (Eq. 3).*

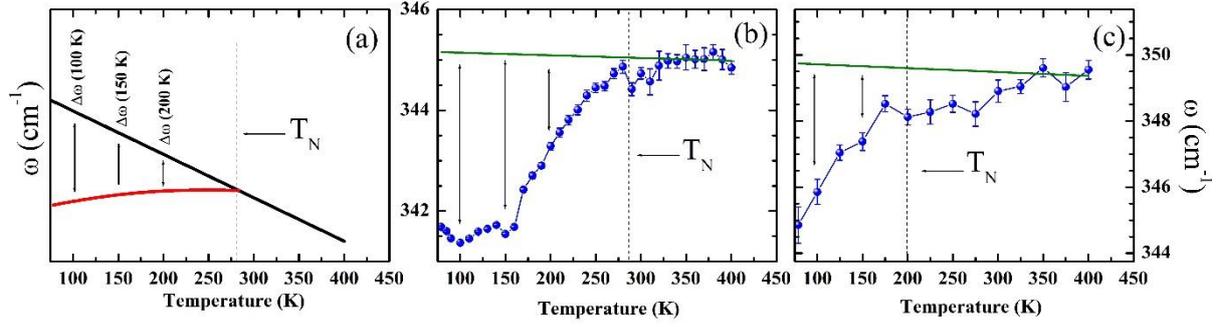

*Figure S3: (a) Schematic diagram to show a deviation in phonon frequency from the expected anharmonic trend due to phonon renormalization below $T_N$. The black line and red line represent the anharmonic trend and renormalized phonon frequency as a function of temperature, respectively. (b) and (c) show the deviations in the experimental values of frequency of the modes P3 ($Ba_{0.9}Sr_{0.1}MnO_3$) and Q3 ($BaMn_{0.9}Ti_{0.1}O_3$) from the anharmonic trends (green lines) below $T_N$.*

## S3. Estimation of spin-phonon coupling

In magnetic materials, the spin-phonon coupling manifests itself by renormalizing the phonon frequency and linewidth (which is related to phonon life-time) in the magnetically ordered phase (below $T_N$). Renormalization of a phonon frequency by the magnetic exchange interaction between spins can be written as:

$$\omega(T) = \omega_{anh}(T) + \lambda <S_i.S_j>$$

thus giving rise to a deviation in the frequency from the expected anharmonic behaviour in the ordered phase by:

$$\Delta\omega(T) = \Delta\omega_{sp-ph}(T) = \omega(T) - \omega_{anh}(T) = \lambda <S_i.S_j>$$

The deviation is shown schematically in Figure S3(a) and for the experimental data in Figure S3(b,c). On the other hand, in the antiferromagnetically ordered phase, the magnetization decreases below $T_N$ with decreasing temperature (see Figure S4(a) for a schematic presentation). In the mean field approximation, considering magnetic interaction only between the nearest neighbour magnetic ions, the deviation in phonon frequency is given as:

$$\Delta\omega_{sp-ph}(T) = \lambda <S_i.S_j> \propto \left[\frac{M(T)}{M_{max}}\right]^2$$

Hence, the deviation in frequency (from the anharmonic behaviour) goes linearly with the square of normalized magnetization (schematic and linear portion of the experimental data for P3 are shown in Figure S4 (b, c)). Slope of this line can be used to estimate the strength of spin-phonon coupling (explained in main text).

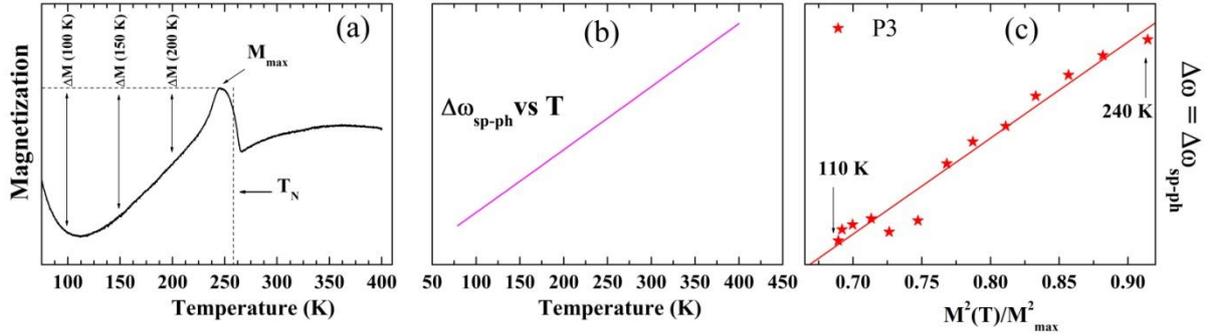

*Figure S4: (a) Magnetization as a function of temperature showing the temperature dependency of $\Delta M = M(T) - M_{max}$. (b) Schematic for the linear relation between deviation of the phonon frequency from the anharmonic behaviour and the square of normalized magnetization, (c ) $\Delta \omega_{sp-ph}(T)$ vs $\left[\frac{M(T)}{M_{max}}\right]^2$ for the P3 phonon.*